\documentclass{natureJLH}
\usepackage[english]{babel}
\usepackage{lineno, blindtext}
\usepackage{epsfig}
\usepackage{subcaption} 
\usepackage{ae, aecompl}
\usepackage{helvet}
\usepackage{epsfig}
\usepackage{dcolumn}
\usepackage{color}
\usepackage[LGR, T1]{fontenc}
\usepackage[latin9]{inputenc}
\usepackage{amssymb}
\usepackage{amsmath}
\usepackage{mwe, tikz}
\usepackage{esint}
\usepackage{longtable}
\usepackage{url}
\usepackage{etoolbox}
\usepackage{multirow}

\newcommand\apj{Astrophys. J.,}
\newcommand\apjl{Astrophys. J. Lett.,}
\newcommand\apjs{Astrophys. J., Suppl.}
\newcommand\mnras{Mon. Not. R. Astron. Soc.,}
\newcommand\nat{Nature,}

\newcommand\nar{New Astron. Rev.}
\newcommand\cjaa{Chin. J. Astron. Astrophys.}
\newcommand\prd{Phys. Rev. D}

\usepackage{xpatch}

\newcommand\xt{\hbox{CDF-S}~XT2}
\newcommand{\src}{XID$_{\rm 7Ms}$~330 }
\newcommand{\aap}{Astron. Astrophys.}
\newcommand{\pasp}{Pub. Astron. Soc. Pacific}

\newcommand{\LuoOneSeven}{{17}}
\newcommand{\XueOneOne}{{38}}
\newcommand{\GuoOneThree}{{22}}
\newcommand{\SantiniOneFive}{{23}}
\newcommand{\HuertasOneFive}{{57}}

\title{A magnetar-powered X-ray transient as aftermath of a binary neutron star merger} 

\author{Y. Q. Xue$^{1,2\star}$,
X. C. Zheng$^{1,2,3\star}$, 
Y. Li$^{4,5}$,
W. N. Brandt$^{6,7,8}$,
B. Zhang$^{9,5,10\star}$,
B. Luo$^{11,12,13}$,
B.-B. Zhang$^{11,12,13}$,
F. E. Bauer$^{14,15,16}$,
H. Sun$^{5}$,
B. D. Lehmer$^{17}$,
X.-F. Wu$^{18,2}$,
G. Yang$^{6,7}$,
X. Kong$^{1,2}$,
J. Y. Li$^{1,2}$,
M. Y. Sun$^{1,2}$,
J.-X. Wang$^{1,2}$, and
F. Vito$^{14,19}$ 
}

\begin{document}

\maketitle

\begin{affiliations}
\item CAS Key Laboratory for Research in Galaxies and Cosmology, Department of Astronomy, University of Science and Technology of China, Hefei 230026, China.
\item School of Astronomy and Space Science, University of Science and Technology of China, Hefei 230026, China.
\item Leiden Observatory, Leiden University, PO Box 9513, NL-2300 RA Leiden, the Netherlands.
\item Kavli Institute for Astronomy and Astrophysics, Peking University, Beijing 100871, China.
\item National Astronomical Observatories, Chinese Academy of Sciences, A20 Datun Road, Beijing 100012, China.
\item Department of Astronomy and Astrophysics, 525 Davey Lab, The Pennsylvania State University, University Park, PA 16802, USA.
\item Institute for Gravitation and the Cosmos, The Pennsylvania State University, University Park, PA 16802, USA.
\item Department of Physics, 104 Davey Lab, The Pennsylvania State University, University Park, PA 16082, USA.
\item Department of Physics and Astronomy, University of Nevada, Las Vegas, NV 89154, USA. 
\item Department of Astronomy, School of Physics, Peking University, Beijing 100871, China.
\item School of Astronomy and Space Science, Nanjing University, Nanjing 210093, China.
\item Key Laboratory of Modern Astronomy and Astrophysics  (Nanjing University), Ministry of Education, Nanjing 210093, China.
\item Collaborative Innovation Center of Modern Astronomy and Space Exploration, Nanjing 210093, China.
\item Instituto de Astrof{\'{\i}}sica and Centro de Astroingenier{\'{\i}}a, Facultad de F{\'{i}}sica, Pontificia Universidad Cat{\'{o}}lica de Chile, Casilla 306, Santiago 22, Chile.
\item Millennium Institute of Astrophysics (MAS), Nuncio Monse{\~{n}}or S{\'{o}}tero Sanz 100, Providencia, Santiago, Chile.
\item Space Science Institute, 4750 Walnut Street, Suite 205, Boulder, Colorado 80301, USA.
\item Department of Physics, University of Arkansas, 226 Physics Building, 825 West Dickson Street, Fayetteville, AR 72701, USA.
\item Purple Mountain Observatory, Chinese Academy of Sciences, Nanjing 210008, China.
\item Chinese Academy of Sciences South America Center for Astronomy, National Astronomical Observatories, CAS, Beijing 100012, China
\\
$^\star$e-mail: xuey@ustc.edu.cn; zheng@strw.leidenuniv.nl; zhang@physics.unlv.edu
\end{affiliations}

%% Abstract
%%
\begin{abstract}
Neutron star-neutron star mergers are known to be associated with short gamma-ray bursts\cite{GW170817,GW170817/GRB170817A,goldstein17,zhangbb18}. If the neutron star equation of state is sufficiently stiff, at least some of such mergers will leave behind a supramassive or even a stable neutron star that spins rapidly with a strong magnetic field (i.e., a magnetar)\cite{dai06,gao06,metzger08,bucciantini12}. Such a magnetar signature may have been observed as the X-ray plateau following a good fraction (up to 50\%) of short gamma-ray bursts\cite{rowlinson13,lv15}, and it has been expected that one may observe short gamma-ray burst-less X-ray transients powered by double neutron star mergers\cite{zhang13,sun17}. A fast X-ray transient (CDF-S XT1) was recently found to be associated with a faint host galaxy whose redshift is unknown\cite{Bauer17}. Its X-ray and host-galaxy properties allow several possible explanations including a short gamma-ray burst seen off axis, a low-luminosity gamma-ray burst at high redshift, or a tidal disruption event involving an intermediate mass black hole and a white dwarf\cite{Bauer17}. Here we report a second X-ray transient, CDF-S XT2, that is associated with a galaxy at redshift $z=0.738$\cite{Zheng17}. The light curve is fully consistent with being powered by a millisecond magnetar. More intriguingly, CDF-S XT2 lies in the outskirts of its star-forming host galaxy with a moderate offset from the galaxy center, as short bursts often do\cite{fong10,berger14}. The estimated event rate density of similar X-ray transients, when corrected to the local value, is consistent with the double neutron star merger rate density inferred from the detection of GW170817\cite{GW170817}. 
\end{abstract}

%% Main text
%%

Upon the completion of the deepest X-ray survey to date, the 7~Ms {Chandra Deep Field-South survey (CDF-S)}, which consists of 102 individual Chandra/Advanced CCD Imaging Spectrometer imaging array (ACIS-I) observations spanning 16.4~yrs\cite{Luo17,Xue17}, we performed a search for X-ray transient events and discovered two notable fast outbursts\cite{Zheng17}, with one dubbed CDF-S XT1 and reported elsewhere\cite{Bauer17} and \xt\ being the focus here.
The Chandra X-ray position of \xt\ is RA$_{\rm J2000.0}$=03$^{\rm h}$32$^{\rm m}$18$^{\rm s}$.38 and DEC$_{\rm J2000.0}$=$-27^\circ$52$^\prime$24$^{\prime\prime}$.2 (with a 1-$\sigma$ positional uncertainty of 0.11$^{\prime\prime}$; see Methods).
This X-ray outburst started at about 07:02:45 Universal Time on 22 March 2015 ($T_0$), and lasted for $\approx20$~ks during an $\approx70$~ks observation (Chandra Observation ID: ObsID 16453).
\xt\ triggered neither of the Gamma-ray Burst Monitor (GBM; 8 keV--30 MeV) and Large Area Telescope (LAT; 20 MeV--300 GeV) on board the Fermi Gamma-ray Space Telescope, the Burst Alert Telescope (BAT) on board the Neil Gehrels Swift Observatory, and the International Gamma-Ray Astrophysics Laboratory/Spectrometer Anticoincidence Shield (INTEGRAL/ACS; 20 keV--8 MeV) (see Methods; and also private communication with A. Lien). It was detected by neither KONUS on board Wind nor the Interplanetary Network (i.e., IPN3) that examines high-energy data from a number of space observatories including, e.g., Fermi, Swift, and INTEGRAL (K.~Hurley and D.~Svinkin, private communication).
Aside from the Chandra observations, no contemporaneous observational data have been identified at any other wavelengths for \xt, spanning $\approx1$~month ahead of the outburst to $\approx4$~months thereafter.

We present the binned Chandra 0.5--7~keV light curves and spectra of \xt\ in Figure~1 for viewing purposes, and fit the unbinned light curves and spectra for physical constraints (see Methods).  
The light curve of \xt\ (listed in Table~1) contains a total of 136 photons, with the $T_{90}$ parameter estimated to be $11.1^{+0.4}_{-0.6}$~ks (i.e., the timespan from the 5\%-th to 95\%-th of the total measured counts; throughout this paper, we quote 1-$\sigma$ errors unless stated otherwise).
The light curve can be well fitted by a broken power-law model using the Markov Chain Monte Carlo code {\tt emcee}\cite{Foreman13}, with the best-fit power-law slopes being $-0.14^{+0.03}_{-0.03}$ before the break (at $t_{\rm b}=2.3^{+0.4}_{-0.3}$~ks) and $-2.16^{+0.26}_{-0.29}$ after the break, respectively (see Fig.~1a). 
We define the hardness ratio HR=(H$-$S)/(H+S), where H and S are the count rates in the 2--7 keV and 0.5--2 keV bands, respectively, and derive its errors based on the Bayesian code {\tt BEHR}\cite{Park06}. 
A simple hardness-ratio analysis reveals an overall softening spectral trend of the source, which is confirmed by a detailed spectral analysis, i.e., the best-fit power-law spectral indexes being $\Gamma=1.57^{+0.55}_{-0.50}$ before the break and $\Gamma=2.53^{+0.74}_{-0.64}$ after the break, respectively (see Figs.~1b, 1c, and Methods).
Given the fact that the light curve of \xt\ peaked quickly 
(with a rest-frame peak luminosity $L_{0.3-10\ \rm keV}\approx3\times 10^{45} \rm\,erg\,s^{-1}$ given our adopted cosmology\cite{Planck16}; see Table~1) 
with a slower decay, we estimate a very short rise time ($\lsim 45$~s) for this outburst (see Methods).

Figure~2a compares the X-ray luminosity light curve of \xt\ with the X-ray afterglow light curves of soft gamma-ray bursts (SGRBs) with known redshifts. One can see that \xt\ is abnormally underluminous compared with SGRB afterglows, especially at early times. Figure~2b presents the isotropic rest-frame $1-10^4$~keV 1-s peak luminosity ($L_{\rm 1-10^4\ keV}$) of SGRB prompt emission against X-ray luminosity at $t=100$~s after the trigger, with \xt\ over-plotted for comparison. The chosen time of $t=100$~s is typical during the X-ray plateau phase\cite{rowlinson13,lv15}. It is clear that if \xt\ originates from the afterglow of an SGRB, at such a low luminosity, the expected $L_{\rm 1-10^4\ keV}$ should be well below the upper limit set by Fermi/GBM. These properties leave open the intriguing possibility that \xt\ could be associated with an undetected low-luminosity SGRB.

The accurate Chandra X-ray position of \xt\ warrants a robust identification of its host galaxy that has a secure spectroscopic redshift of $z=0.738$ and an apparent AB magnitude of $m_{\rm F160W}\approx24$~mag\cite{Guo13,Santini15}, with an offset of $0.44^{\prime\prime}\pm 0.25^{\prime\prime}$ (i.e., a projected distance of $3.3\pm1.9$~kpc; the error is computed as the root of quadratic sum of the Chandra and HST positional uncertainties as well as the uncertainty of astrometric registration between these two sets of data) considering the peak-flux position of the host galaxy or $0.45^{\prime\prime}\pm 0.25^{\prime\prime}$ ($3.3\pm1.8$~kpc) considering the {\tt SExtractor}-derived position (see Fig.~3a and Methods). 
Based on the galaxy surface density derived from the HST/CANDELS F160W DR1 catalog\cite{Guo13}, we estimate that the probability of a coincident match between \xt\ and a galaxy brighter than $m_{\rm F160W}\approx24$~mag within $0.44''$ is only $\approx1$\%.
We adopt the median stellar mass ($M_*=1.17\times10^9\ M_\odot$), star-formation rate (SFR$=0.81\ M_\odot$~yr$^{-1}$), and metallicity ($Z=2.0\ Z_\odot$) of the host galaxy from the five independent and consistent estimates derived from spectral energy distribution (SED) fitting\cite{Santini15}, which utilized the same photometry, galaxy templates, and initial mass function\cite{BC03} as well as different star-formation history models and extinction laws.  
Given the values of redshift, stellar mass, and SFR, the host galaxy is located within the lower part of the galaxy main sequence\cite{Speagle14}, i.e., having a relatively low SFR given its stellar mass and sitting close to the lower bound of the main sequence.

We present additional host-galaxy related properties of \xt\ in Fig.~3. 
Using the HST F125W-band image, we compute the offset between \xt\ and the host galaxy in units of galaxy half-light radius ($R_{50}=0.38''$), and find that it is well within the distribution of known SGRB-host galaxy offsets\cite{fong10,berger14} (see Fig.~3b).
We also calculate the light fraction $F_{\rm light}$ that indicates how bright the \xt\ region (i.e., the red circle with $r=0.11^{\prime\prime}$ in Fig.~3a) is relative to the other parts of the host galaxy, with $F_{\rm light}$=1 (0) standing for the brightest (faintest) region.
We utilize the segmentation given by {\tt SExtractor} to define the host-galaxy region,
and compute $F_{\rm light}$ as the ratio of total light of 
the galaxy region with surface brightness smaller than the median value within the \xt\ region 
to the entire galaxy region.
We obtain $F_{\rm light}$=0.0 for \xt, which is consistent with $F_{\rm light}$ values of the majority of known SGRBs (see Fig.~3c).

To discern better whether \xt\ has an NS-NS merger origin, we calculate the probability, O(II:I)$_{\rm host}$, of the source being similar to long GRB (LGRB: massive star collapse type or Type II) versus SGRB (compact star merger type or Type I) populations based on the statistical properties of the host galaxy data of the two types\cite{fong10,blandchard15,li16}. 
The criteria used include how each of the following observed parameters compares with the distributions of both LGRBs and SGRBs collected\cite{li16}: stellar mass, SFR, metallicity, offset, and galaxy size.
The probability for each criterion for each category is calculated, and O(II:I)$_{\rm host}$ is defined as the product of the long-to-short GRB probability ratios for all the criteria.
By definition, a negative (positive) log[O(II:I)$_{\rm host}$] value indicates a merger (collapsar) origin.
We obtain log[O(II:I)$_{\rm host}$]=$-$0.8, which is roughly the median value of known SGRBs and is smaller than that of 98\% of known LGRBs (see Fig.~3d).
This indicates that \xt\ is very likely of a compact star merger origin.

A rapidly spinning magnetar has a spindown luminosity that evolves with time as $L_{\rm sd} \propto L_{0}/(1+t/t_{\rm sd})^2$. 
This can be approximated as $L_{\rm sd} \propto t^0$ for $t \ll t_{\rm sd}$ and $L_{\rm sd} \propto t^{-2}$ for $t \gg t_{\rm sd}$\cite{Dai98,zhang01}. 
The observed light curve is consistent with such an evolution (see Fig.~1a and Methods). 
At $z=0.738$, an SGRB with $L_{1-10^4 \ {\rm keV}} < 1.5 \times 10^{51} \ {\rm erg \ s^{-1}}$ (including 170817A-like GRBs) would be too faint to trigger Fermi/GBM and other GRB detectors (see Methods). 
Therefore, \xt\ could be, in principle, associated with a low-luminosity SGRB below the Fermi and INTEGRAL detection limits. 
In any case, the lack of a detectable SGRB is consistent with an off-axis jet configuration. 
Such a geometry has a larger probability to be detected, consistent with the possibility that \xt\ is the first such event detected.

We estimate the event rate density (or the volumetric rate) of \xt-like events to be 1.3$_{-1.1}^{+2.8}\times10^4$~Gpc$^{-3}$~ yr$^{-1}$, taking into account a number of factors that include, e.g., the event searching procedure, varying sensitivities across the Chandra/ACIS-I field of view (FOV), and the X-ray spectral shape and peak luminosity
of \xt\ (see Methods).
Note that CDF-S XT1 is not included to estimate the rate, since its observational properties are different from \xt\ and it likely belongs to a different type of transients (see Methods).
Based on the redshift evolution of the event rate density of SGRBs given three different merger delay models\cite{Sun15}, one can derive the corresponding local event rate density of \xt-like events, which is 1.8$_{-1.6}^{+4.1}\times10^3$~Gpc$^{-3}$~ yr$^{-1}$. 
This is consistent with 
the NS-NS merger event rate density inferred from the detection of GW170817 by the LIGO-Virgo Collaboration\cite{GW170817}, i.e. 1.5$_{-1.2}^{+3.2}\times10^3$~Gpc$^{-3}$~ yr$^{-1}$. 
The lower limit on the rate density remains consistent with that of 170817A-like SGRBs (i.e., $190^{+440}_{-160}$~Gpc$^{-3}$~ yr$^{-1}$) inferred from the Fermi/GBM detection\cite{zhangbb18}. This leaves open the possibility that the viewing angle of \xt\ could be  comparable to or even smaller than that of GW170817/GRB 170817A, even if the viewing angle of the former is likely larger.

The Thomson optical depth for X-rays should be below unity so that they can escape from the surrounding ejecta from the merger.
This requires that the viewing angle is in the so-called ``free zone''\cite{sun17} if the merger occurred right before the onset of X-ray emission, which demands that the line of sight is already cleared by a low-luminosity jet-like outflow\cite{metzger08,bucciantini12}. Numerical simulations\cite{bucciantini12} showed that a magnetar can open a funnel with a moderate opening angle of $\sim 1$ radian within $\sim$ 100 s, and a possible undetected low-luminosity SGRB could also help open the gap. The required magnetar parameters can be made consistent with those inferred from the observations (see Methods).

Other mechanisms to produce cosmological X-ray transients are disfavored by the observational data of \xt\ (see Methods). Low-luminosity LGRBs or massive-star shock breakout events are typically associated with active star formation. This is inconsistent with the host-galaxy type of \xt\ and its large offset with respect to the host-galaxy center. GRB orphan afterglows and tidal disruption events typically have much longer durations and very different light-curve shapes. Comparison of \xt\ with future multi-messenger observations of X-ray transients directly detectable by gravitational wave detectors can help verify its NS-NS merger origin, and help provide unprecedented insight into the physics of NS-NS mergers.

%% Methods
%%

\begin{figure}[t]
\begin{tabular}{c}
\includegraphics[keepaspectratio, clip, width=1.0\textwidth]{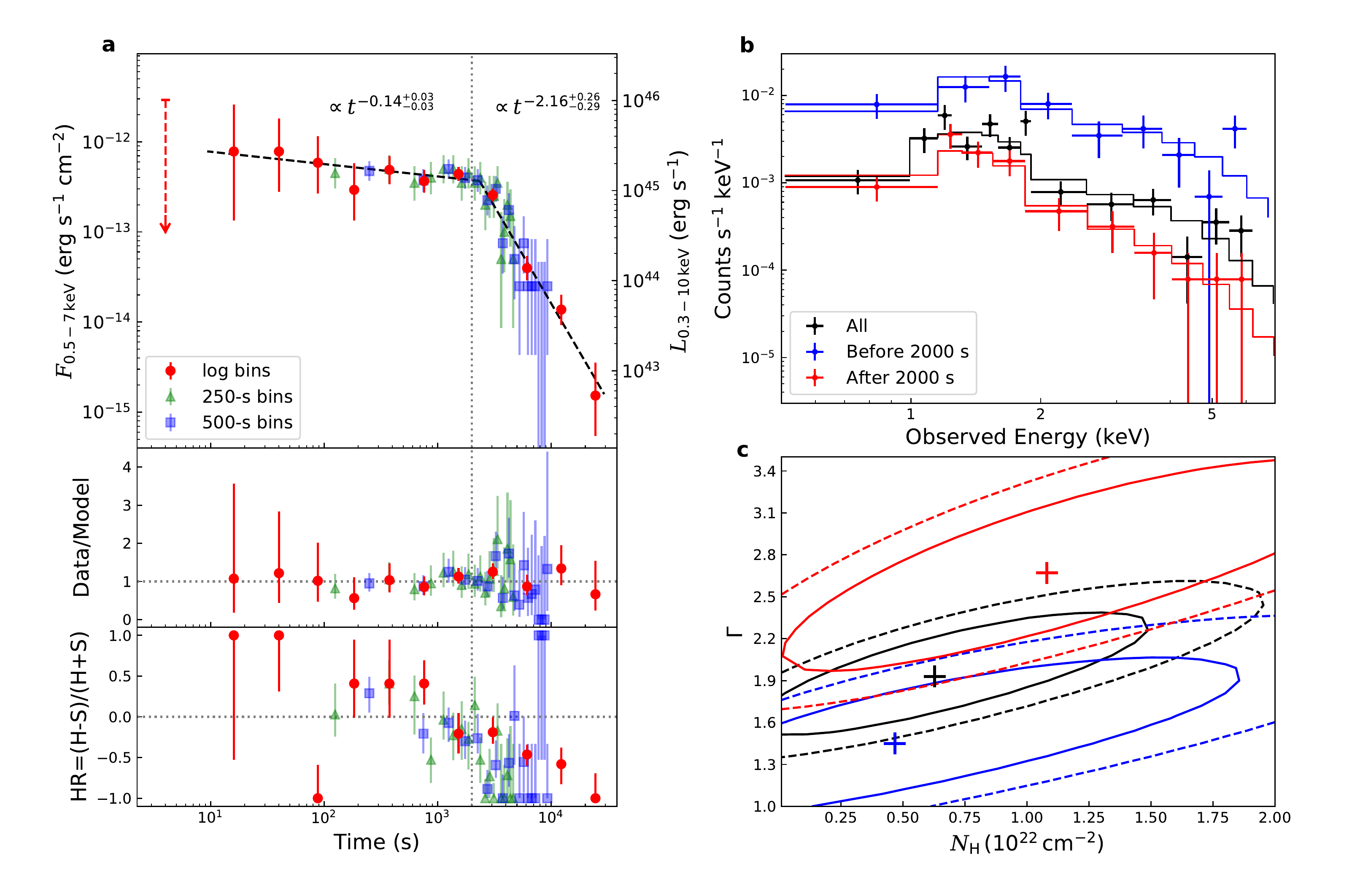} \\
\end{tabular}
\caption{{\bf Timing and spectral evolution of \xt. a,} Light curves (see Table~1 for additional information) and best-fit broken power-law model, ratio between data and best-fit model, and hardness-ratio evolution of \xt.
The downward dashed arrow indicates no photons being detected from \xt\ at $t<10$~s, and
the corresponding 1-$\sigma$ flux upper limit is $3.1\times10^{-16}$~erg~cm$^{-2}$~s$^{-1}$ (see Methods).
The vertical dotted line ($t$=2000~s) roughly indicates the break time that divides the outburst into two segments. Three different binning schemes are presented to show the details around the break. {\bf b,} Spectra and best-fit models of the entire outburst and the two segments.
1-$\sigma$ errors are plotted in both {\bf a} and {\bf b}.
{\bf c,} Corresponding best-fit values of photon index $\Gamma$ and hydrogen column density $N_{\rm H}$ as well as respective 1-$\sigma$ (solid curves) and 2-$\sigma$ (dashed curves) confidence contours.
}
\label{fig:lc_sp}
\end{figure}

\begin{figure}[t]
\begin{tabular}{c}
\includegraphics[keepaspectratio, clip, width=1.0\textwidth]{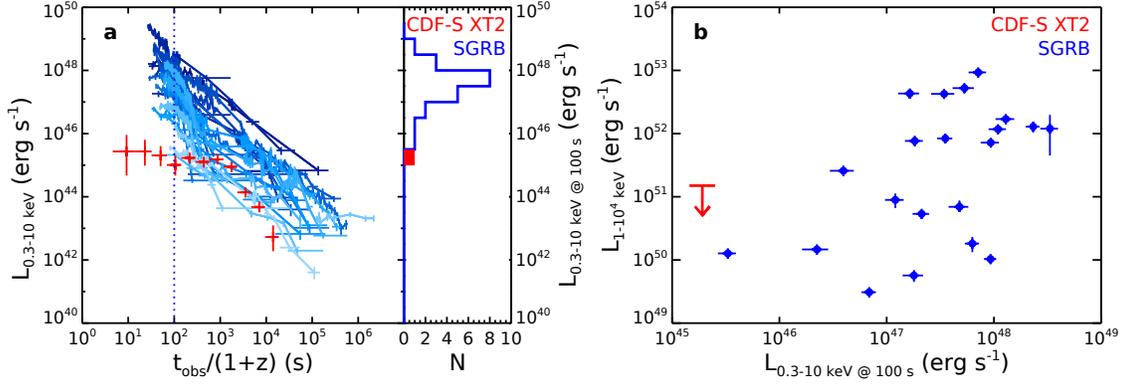} \\
\end{tabular}
\caption{{\bf X-ray and gamma-ray luminosity related information of \xt. a,} X-ray luminosity light curve and X-ray luminosity at $t=100$~s of \xt~(in red), in comparison with that of known SGRB X-ray afterglows (in blue colors) that have redshift information. {\bf b,} Isotropic rest-frame \hbox{1--10$^4$~keV} 1-s peak luminosities of SGRB prompt emission versus X-ray luminosities at $t=100$~s after the trigger, with \xt\ over-plotted for comparison. Errors and upper limits are quoted at 1-$\sigma$ level.
}
\label{fig:host}
\end{figure}

\begin{figure}[t]
\begin{tabular}{c}
\includegraphics[keepaspectratio, clip, width=1.0\textwidth]{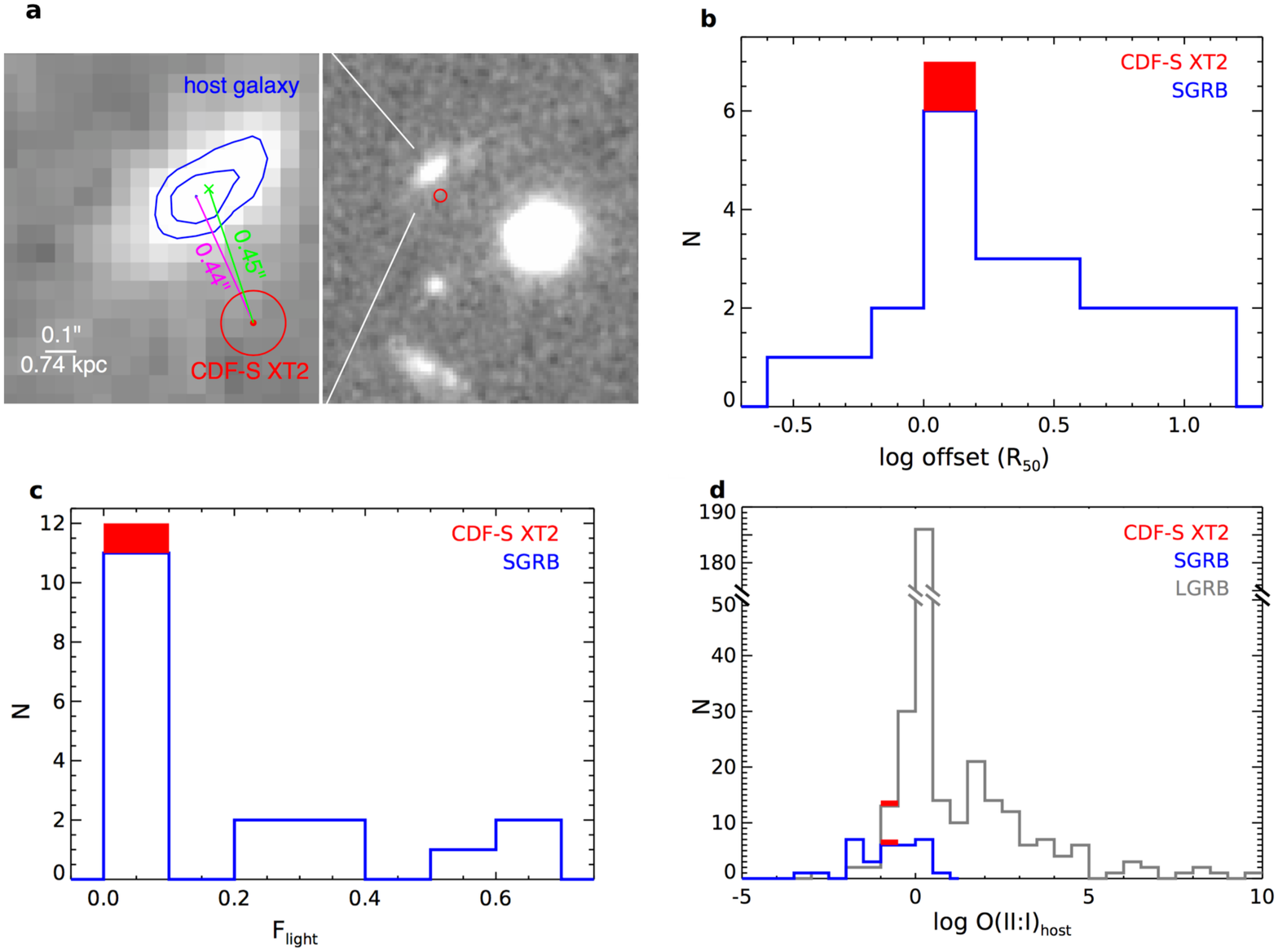} \\
\end{tabular}
\caption{{\bf Host-galaxy related properties of \xt. a,} HST/CANDELS F125W-band image and zoomed-in portion of the host galaxy showing the positional offset between \xt\ (encircled by the red $r=0.11^{\prime\prime}$ circle) and its host galaxy (contours shown atop). {\bf b,} Histogram of GRB-host galaxy offsets (in units of $R_{50}$) for known SGRBs. {\bf c,} Histogram of $F_{\rm light}$ for known SGRBs. {\bf d,} Histogram of log[O(II:I)$_{\rm host}$] for known SGRBs and LGRBs.
\xt\ is overlaid in red in panels (b--d), which is consistent with being a compact star merger origin.
}
\label{fig:lc_sp}
\end{figure}

\clearpage
\begin{methods}

\subsection{Chandra data reduction and extraction of light curves and spectra.}

We reduce and analyze the 7~Ms \hbox{CDF-S} observations\cite{Luo17} using the Chandra Interactive Analysis of Observations (CIAO; v4.8) tools, MARX ray-tracing simulator (v5.3), ACIS Extract (AE; version 2016may25)\cite{Broos10}, and custom software.
Briefly, we apply charge transfer inefficiency corrections, remove bad pixels, flag cosmic-ray background events, discard faint afterglow events with $\ge3$ total counts that fall onto the same pixel within 20~s, and reject background flares in an appropriate way\cite{Luo17} to obtain a cleaned event file for each individual observation. 
We then register and align all individual observations to a common accurate astrometric frame (i.e., the TENIS $K_s$-band catalog\cite{Hsieh12}) and merge them into a combined master event file that is ready for the exaction of images, light curves, spectra, etc.
Our procedure of data reduction ensures the best possible X-ray source positions and reliable photometry, which is particularly critical for faint sources.

Using the cleaned master event file, we extract the 0.5--7~keV light curves and spectra of \xt\ (see Figure~1) within a $r=3.5^{\prime\prime}$ (corresponding to an encircled energy fraction of $\approx95$\% given its off-axis angle of 4.5$^\prime$ in ObsID 16453) circular source region centered at its position. Its photons were detected only amid ObsID 16453 and aggregated 136 in total (see Table~1), which enable AE to derive accurate X-ray position of the source by taking the local ACIS-I point spread function (PSF) into account.
The source position and associated uncertainty are given by the AE keywords of RA\_DATA, DEC\_DATA, and ERR\_DATA, 
which are the mean/centroid data position and corresponding standard error that is  
computed using the variances of the PSF and flat background within the extraction region, respectively.
Over the duration of \xt, the individual pixels that recorded the detected photons traced out portions of the Lissajous pattern expected due to Chandra dither, indicating that the source is indeed celestial.
The background level is very low in the source region, indicating a highly significant detection: we extract the background in a source-free annulus centered around the source whose area is 10 times that of the source region, and find only 7 background photons during the outburst, which indicates 0.7 photons expected in the source region.
This background level is consistent with the mean background level of 0.184 photons $\rm Ms^{-1}\,pixel^{-1}$ in the 7~Ms CDF-S\cite{Luo17}, and there is no sign of a peak of the background flux throughout the outburst.
Hence we conclude that we can ignore the influence of the background in our analysis.

\subsection{No high-energy (gamma-ray) trigger.} CDF-S XT2 was in the FOV of Fermi/GBM during $T_0\pm1000$~s. We examine the light curves of the 8 GBM detectors around $T_0$, whose pointing angles were within 60 degrees with respect to the source location. We find no significant source-like gamma-ray emission signal above background. 
We then extract the spectra of GBM detectors n4, n5, and b0 during $T_0\pm25$~s, confirming that the extracted spectra are consistent with the background spectra.
Subsequently, we calculate the source count limits\cite{Kraft91} at 90\% confidence level (S$_{LL,i}$, S$_{UL,i}$) in each energy channel $i$ based on the corresponding observed counts and background counts, and obtain the flux upper limits by fitting the power-law model to a simulated spectrum realized based on (S$_{LL,i}$, S$_{UL,i}$).
The flux upper limits are $f_{\rm 1-10^4\ keV}=6.0^{+0.7}_{-0.7}\times10^{-7}$ erg~cm$^{-2}$~s$^{-1}$, $f_{\rm 0.3-30\ keV}=2.4^{+5.3}_{-2.1}\times10^{-9}$ erg~cm$^{-2}$~s$^{-1}$, and $f_{\rm 8-100\ keV}=1.4^{+0.3}_{-0.3}\times10^{-8}$  erg~cm$^{-2}$~s$^{-1}$, respectively, with the corresponding isotropic rest-frame luminosity upper limits being
$L_{\rm 1-10^4\ keV}=1.5\times10^{51}$ erg~s$^{-1}$, $L_{\rm 0.3-30\ keV}=6.1\times10^{48}$ erg~s$^{-1}$, and $L_{\rm 8-100\ keV}=3.5\times10^{49}$ erg~s$^{-1}$, respectively. 
We also estimate Fermi/LAT flux and isotropic rest-frame luminosity upper limits, during $T_0\pm10^5$~s, to be 
$f_{\rm 100\ MeV-30\ GeV}\approx6.0\times10^{-10}$ erg~cm$^{-2}$~s$^{-1}$ and $L_{\rm 100\ MeV-30\ GeV}\approx1.5\times10^{48}$ erg~s$^{-1}$, respectively.

A search for a gamma-ray component temporally and spatially coincident with \xt\ with the Swift/BAT data also led to a negative result (A. Lien, private communication). This is consistent with the facts that Fermi/GBM is more sensitive than Swift/BAT in detecting SGRBs and that the BAT SGRB population is consistent with the GBM SGRB population\cite{burns16}.
%the aforementioned GBM upper limits can be safely adopted for the putative SGRB.

\subsection{X-ray spectral fitting.} 
To inspect the spectral variation, we not only use {\tt XSPEC}\cite{Arnaud1996} to fit the entire unbinned spectrum (Spec\_0) throughout the event, but also fit the unbinned spectra before (Spec\_1) and after 2000~s (Spec\_2) with the Cash statistic ($C$), which is the dividing point close to the break time (see Fig.~1a) and can balance the total counts in the two segments.
The model we utilize is {$\tt phabs\times(zphabs\times zpow)$}, which includes the Galactic absorption (fixed to a column density of $8.8\times 10^{19}$~cm$^{-2}$)\cite{Stark1992}, intrinsic absorption ($N_{\rm H}$), and intrinsic power-law component ($\Gamma$), and fits all the spectra well (see Table~2 for details).
We obtain $\Gamma =1.93_{-0.47}^{+0.52}$ and $N_{\rm H}=0.61_{-0.61}^{+1.00}\times 10^{22} \rm\,cm^{-2}$ for Spec\_0 (see Fig.~1c), and derive an isotropic equivalent emission energy in the 0.3--10~keV band of $E_{\rm iso, 0.3-10\,keV} \approx 2.4\times 10^{48}$~erg for \xt.
When fitting Spec\_1 and Spec\_2 jointly, we consider four cases, i.e., Case~A (free $\Gamma$ and free $N_{\rm H}$; see Fig.~1c), Case~B (free $\Gamma$ and linked $N_{\rm H}$), Case~C (linked $\Gamma$ and free $N_{\rm H}$), and Case~D (linked $\Gamma$ and linked $N_{\rm H}$), and adopt the Akaike information criterion\cite{Akaike74} (AIC=$C+2k$, where $k$ is the number of free parameters in the model) to identify which model fits the data best.

According to Table~2, Case~B has the smallest AIC and therefore describes Spec\_1 and Spec\_2 best, indicating that $N_{\rm H}$ is most likely constant throughout the outburst and $\Gamma$ increases from Spec\_1 to Spec\_2 (i.e., spectral softening, which is significant at a confidence level of $\approx 89$\% given exp((AIC$_{\rm B}-$AIC$_{\rm D})/2)\approx0.11$; also see Fig.~1c).  
This likely spectral evolution cannot be compared with the magnetar model predictions
because the latter are currently unavailable.

\subsection{Estimation of rise time.}
In Fig.~1a, it is shown that the flux reaches its peak at the very beginning, which implies an extremely short rise time. 
However, because of the small number of counts in the first few bins, we cannot tell the exact position of the peak easily. 
There are two possibilities: the first photon is in the rising period of the light curve, hence the high flux is just due to our binning strategy; or, it is really at the peak.
If the first scenario is true, we should find some clues in the analysis of the intervals between the recorded arrival times of the photons during the beginning of the light curve, but we find that the intervals between the first few photons do not show any particular pattern.
We also inspect the relative positions of the first 19 photons arriving within the first six bins in the event (see Table~1) and find that each pair of neighboring photons is well separated, which excludes the possibility of any residual cosmic-ray effect during the period. 
We take this as evidence that the first photon is at the peak or at least near the peak.

In the case of not detecting any photons in the rising period, we can estimate the rise time based on the Poisson distribution. 
Assuming that the rise profile is linear and the rise time is $T_{\rm r}$, we can write the probability that we do not observe any photon at the $i$th frame since the event occurs,
\begin{equation}
    p_{T_{\rm r},i}(0)=e^{-\frac{f_{\rm m}}{T_{\rm r}}i t_{\rm f}^2},
\end{equation}
where $f_{\rm m}$ is the maximum photon flux ($\approx 0.05$~counts~s$^{-1}$) of the event and $t_{\rm f}\approx3.2$~s is the ACIS-I frame time. We also assume $T_{\rm r}\approx n t_{\rm f}$.
Then the probability of not detecting any photon during the whole rising period and the probability distribution of $T_{\rm r}$ should be 
\begin{eqnarray}
    p_{T_{\rm r}}(0)&=&\prod_{i}^n p_{T_{\rm r},i}(0)=e^{-\frac{f_{\rm m}T_{\rm r}}{2}}\approx e^{-\frac{f_{\rm m}}{2}n t_{\rm f}}, {\rm and}\\
    P(n)&=&\frac{e^{-\frac{f_{\rm m}}{2}n t_{\rm f}}}{\sum\limits_{i=0}^{\infty}e^{-\frac{f_{\rm m}}{2}i t_{\rm f}}}=\frac{f_{\rm m}}{2}e^{\frac{-n t_{\rm f}f_{\rm m}}{2}}=\frac{f_{\rm m}}{2}e^{\frac{-T_{\rm r}f_{\rm m}}{2}}.
\end{eqnarray}
Based on the above deduction, we can obtain the 1~$\sigma$ upper limit on the rise time, $\approx45$~s (i.e., $\approx26$~s in the rest frame).

During the period from the start time of ObsID 16453 to the arrival of the first photon of \xt, we estimate a background photon flux level of $(2.4\pm0.6)\times10^{-5}$~counts~s$^{-1}$, which corresponds to a flux upper limit of $(3.1\pm0.8)\times10^{-16}$~erg~cm$^{-2}$~s$^{-1}$ (assuming $\Gamma=1.4$ that is the spectral slope of the cosmic X-ray background) before the onset of the outburst.

\subsection{Determination of the offset between \xt\ and its host galaxy.}
There are 18 HST/WFC3 F125W-band exposures that cover \xt\ during 12 visits.
For each observation, we only consider the portion of the image that is local to \xt, i.e., within a roughly circular area with $r\approx2^\prime$ around \xt, in order to reduce the likely effect of astrometric variation across the FOV.
We utilize {\tt SExtractor} to find sources in the 18 local images that are free of cosmic-ray events, and then register and combine them together using the standard commands {\tt tweakreg} and {\tt AstroDrizzle}, respectively.
The root mean squares of the registrations between different images are $\approx0.1$~pixel (with a pixel size of 0.06$^{\prime\prime}$), which indicates good astrometric registrations. 
Subsequently, we use {\tt SExtractor} to find sources in the local combined F125W image and register it to the astrometric frame of the 7~Ms CDF-S main catalog\cite{Luo17}. 
By doing this, we ensure that the X-ray image of \xt\ from ObsID 16453 and the local combined F125W image have the same astrometric frame (accurate to $\approx 0.2^{\prime\prime}$), which warrants a reliable determination of the offset between \xt\ and its host galaxy (see Fig.~3a).   

We also perform the above procedures using the 16 HST/WFC3 F160W-band observations that cover CDF-S XT2 and obtain essentially the same results.
Here we choose to report the F125W-band results, 
given that the F125W band probes a median rest-frame wavelength $\approx 7200$~\AA\ and trace the stellar distribution of the galaxy better than the F160W band (corresponding to $\approx 9200$~\AA).  
\subsection{Multiwavelength observations in the \xt\ neighborhood.} 
We display in Fig.~4 the 0.5--7~keV\cite{Luo17} and HST/CANDELS F160W-band\cite{Guo13} images of the \xt\ neighborhood, and show in Table~3 the properties of the 6 closest galaxies\cite{Santini15} within a radius of 5$^{\prime\prime}$ around \xt\ in the HST/CANDELS F160W DR1 catalog\cite{Guo13}.

The X-ray position of \xt\ (with a total of 136~photons detected) is solely determined by ObsID 16453 within which its outburst occurred, which is slightly ($\approx0.3^{\prime\prime}$, being smaller than the X-ray image pixel size of 0.492$^{\prime\prime}$) northeastward of the X-ray position of the source \src reported in the 7~Ms CDF-S main catalog\cite{Luo17}.
\src was initially identified as being the same source as XID$_{\rm 4Ms}$~256 in the 4~Ms CDF-S main catalog\cite{Xue11}.
XID$_{\rm 4Ms}$~256 had only $\approx30$ photons detected during the 4~Ms exposure, and was classified as a normal galaxy. 
%\\
When the 7~Ms survey was completed, \src was reported with a total of over 170 photons detected and was classified as an active galactic nucleus.
Only upon further investigation of the X-ray variability\cite{Zheng17} was it realized that the flux and position of \src are actually a composite of two independent sources --- XID$_{\rm 4Ms}$~256 and \xt, which is clearly shown in Fig.~4 and Table~3.

The host galaxy of \xt\ is a dwarf galaxy (labeled as \#1) with a secure spectroscopic redshift of 0.738 and a morphology of irregular disk, while the X-ray photons of XID$_{\rm 4Ms}$~256 are from an elliptical galaxy (labeled as \#2) with a secure spectroscopic redshift of 0.740 and a morphology of pure bulge.
These two galaxies likely belong to the same prominent large-scale structure at $z\approx0.73$ in the Extended-CDF-S\cite{Silverman10,Xue17}.
Given that XID$_{\rm 4Ms}$~256 has a low count rate of $7.5\times 10^{-6}$~counts\,s$^{-1}$, similar to the average level of \src without the contribution of ObsID\,16453, the expected contribution from the elliptical galaxy during the outburst of \xt\ is $\lsim0.2$~photons, and can be ignored when analyzing Chandra data of \xt.

\subsection{Magnetar parameters.}
Assuming dipolar spindown, the magnetar parameters may be estimated using the relations\cite{zhang01}:
\begin{eqnarray}
 L_{\rm sd} & \simeq & (10^{49} \ {\rm erg \ s^{-1}}) B_{p,15}^2 P_{0,-3}^{-4} R_6^6, {\rm and }\\
 t_{\rm sd} & \simeq & (2050\ {\rm s}) I_{45} B_{p,15}^{-2} P_{0,-3}^2 R_6^{-6},
\end{eqnarray}
where $B_p = (10^{15}\ {\rm G}) B_{p,15}$ is the surface magnetic field at the pole, $P_0 = (10^{-3}\ {\rm s}) P_{0,-3}$ is the initial spin period, $I = (10^{45}\ {\rm g \ cm^2}) I_{45}$ is the moment of inertia of the NS, and $R = (10^6\ {\rm cm}) R_6$ is the NS radius. Assuming $L_x = \eta L_{\rm sd}$, where $\eta$ is the efficiency of converting spindown luminosity to X-ray luminosity ($\eta=10^{-3}\eta_{-3}$), and noticing that $I_{45} \sim 1.9$ for a supramassive NS after the merger of two NSs, one obtains
\begin{eqnarray}
B_{p,15}\simeq 1.6 \eta_{-3}^{1/2} (\frac{L_{\rm X}}{3\times 10^{45}\ {\rm erg\ s}^{-1}})^{-1/2} (\frac{t_{\rm sd}}{2.3\ \rm ks})^{-1} I_{45} R_{6}^3, {\rm and}
\\
P_{0,-3} \simeq 1.7 \eta_{-3}^{1/2} (\frac{L_{\rm X}}{3\times 10^{45}\ {\rm erg\ s}^{-1}})^{-1/2} (\frac{t_{\rm sd}}{2.3\ {\rm ks}})^{-1/2} I_{45}^{1/2},
\end{eqnarray}
if one scales $L_{\rm X}$ to the \xt\ peak luminosity and assumes $t_{\rm sd} = t_b = 2.3^{+0.4}_{-0.3}$ ks. One can see that reasonable magnetar parameters can be obtained if $\eta$ is of the order of $10^{-3}$. 
With such parameters and assuming that the ejecta mass is $M_{\rm ej} \sim (0.01\ M_\odot) M_{\rm ej,-2}$, one has a magnetar-ejecta interaction parameter $\dot E/M_{\rm ej} = 3\times 10^{50} \ {\rm erg \ s^{-1} \ M_\odot^{-1}} \eta_{-3}^{-1} M_{\rm ej,-2}^{-1}$, which is between the ``low'' and ``average'' cases studied\cite{bucciantini12} (see their Fig.~2). Given such parameters, the magnetar wind could open a reasonably large funnel within $\sim$ 100 s, so that one can detect X-rays at a relatively large viewing angle, as might be the case for \xt.

\subsection{Estimation of event rate density.}
While searching for transient events in the 7~Ms \hbox{CDF-S}\cite{Zheng17}, we required that the sources were covered by all the 102 CDF-S observations, which limits the actual FOV considered to the central $r=8^{\prime}$ area of the CDF-S survey.
Since the detection limit of the 7~Ms CDF-S is a function of off-axis angle, we divide the central $r=8^{\prime}$ FOV into a series of narrow concentric annuli with a width of $\Delta r$ to determine the minimum 0.5--7~keV counts of an X-ray source required for a detection in each annulus, i.e., the detection limit in each annular region.
Following the original transient-searching procedure\cite{Zheng17} and considering that the background of the 7~Ms CDF-S is stable\cite{Luo17},  
we assume a background region 10 times that of the source region and estimate the expected background counts for the source in a 70~ks observation (i.e., the exposure of ObsID 16453) using the mean background count rate of the 7~Ms CDF-S\cite{Luo17}. 
The fluctuation ($\sigma$) of the background counts is given by the Poisson distribution. 
For short events like \xt, we assume that all the photons during the outburst can be caught in a single Chandra observation of typical exposure. 
If the net counts of the source exceed the mean background level by 3~$\sigma$, then we  
consider it as a detected X-ray transient.
For events with a similar spectral shape and peak luminosity to that of \xt, we can then utilize the 3~$\sigma$ counts limit to derive the maximum redshift $z_{\rm m}(r)$ at which we may detect such an event.
Combining the FOV size being considered and the rest-frame monitoring time, finally, we estimate the observed event rate density following 
\begin{equation}    
    \langle\dot{N}\rangle = 1/\int_{0}^{8^{\prime}}2\pi r{\rm d}r\int_{0}^{D_c(z_{\rm m}(r))}D_c(z)^2\frac{7\,{\rm Ms}}{1+z}\,{\rm d}D_c(z),
\end{equation}
where $D_c(z)$ denotes the comoving distance at redshift $z$.
This yields an event rate density of $1.3^{+2.8}_{-1.1}\times 10^4$~Gpc$^{-3}$~yr$^{-1}$ at an average $z_{\rm m}$=1.9 (with $z_{\rm m}(0)$=2.1 and $z_{\rm m}(8)$=1.7), which can then be converted into a local value 
(i.e., $1.8^{+4.1}_{-1.6} \times 10^3$ ~Gpc$^{-3}$~yr$^{-1}$) appropriately\cite{Sun15}.

This event rate density is consistent with the NS-NS merger event rate density, $1.5^{+3.2}_{-1.2} \times 10^3$ ~Gpc$^{-3}$~yr$^{-1}$, inferred from the discovery of GW170817\cite{GW170817}. The case of a binary NS-NS merger origin for \xt\ requires that a significant fraction of NS-NS mergers leave behind a long-lived massive NS remnant. The fraction depends on the unknown NS equation of state. The case of GW170817/GRB 170817A is inconclusive. Even though observations are consistent with having a black hole merger product\cite{margalit17,pooley18,ruiz18,rezzolla18}, the existence of a long-lived NS engine is not ruled out and could be helpful in interpreting some phenomena of the event\cite{ai18,yu18,li18,piro19}. Interestingly, in order to interpret the internal X-ray plateau data of SGRBs, a relatively large maximum NS mass is needed, so that a significant fraction of NS-NS mergers would leave behind long-lived NSs\cite{lasky14,gao16,li16b}.

\subsection{Comparison between CDF-S XT1 and CDF-S XT2.}

CDF-S XT1 was another X-ray transient discovered from CDF-S\cite{Bauer17}. Since there is no spectroscopic-redshift measurement of its faint host galaxy, one cannot reliably compare the luminosities and other redshift-dependent properties of the two events. In any case, the observed properties are already different from that of CDF-S XT2. 
First, its light curve is characterised with a rapid rise followed by an immediate decay with a slope of $-1.53\pm 0.27$. This is in stark contrast to the plateau behavior and the subsequent steeper decay as displayed in CDF-S XT2. While the light curve of CDF-S XT2 shows an ironclad signature of magnetar emission, that of CDF-S XT1 is difficult to reconcile within the magnetar model. Second, without a proper redshift measurement, one cannot conduct the same host-galaxy related analysis as done here for CDF-S XT2, which leads us to conclude that its host-galaxy properties are much more consistent with SGRBs rather than LGRBs. 
Indeed, besides NS mergers, several other interpretations are allowed for \hbox{CDF-S~XT1}, including an ``orphan'' X-ray afterglow from an off-axis SGRB with weak optical emission, a high-redshift low-luminosity GRB without prompt emission below rest-frame $\sim$20~keV, or a highly beamed tidal disruption event (TDE) involving an intermediate-mass black hole and a white dwarf with little variability\cite{Bauer17}. For the above reasons, we do not conclude that CDF-S XT1 shares the same origin as CDF-S XT2 (even though this possibility is not ruled out), and do not include CDF-S XT1 in estimating the event rate density of CDF-S XT2-like transients.

\subsection{Inconsistency of CDF-S XT2 properties with X-ray transients of other origins.}
We discuss other X-ray transient types that might be considered to interpret \xt\ and explain why they are inconsistent with the data.

\begin{itemize} 
 \item X-ray selected TDEs\cite{komossa15} typically have a much longer duration than \xt. They tend to be located at the centers of host galaxies where the supermassive black holes reside. Both the light-curve shape and the large offset with respect to the host-galaxy center of \xt\ generally disfavor the TDE origin. 
 Some special types of TDEs, such as white dwarf-intermediate mass black hole (IMBH) TDEs may have shorter durations, but the peculiar plateau and the post-plateau rapid decay of \xt\ cannot be interpreted with the TDE model. Indeed, the jetted TDE source, Sw 1644+57\cite{burrows11}, which has been interpreted as one such TDE, shows very different properties, including a long ($>10$~day) extended light curve. Furthermore, in view that no IMBHs have been identified for certain, the event rate density of such systems is difficult to estimate, and may not reach the high value as inferred for \xt.
 \item An orphan GRB afterglow has a light curve characterized by a rapid rise followed by a steep decay\cite{granot02}. No X-ray plateau is expected. The shape of the \xt\ light curve disfavors an orphan afterglow origin.
 \item Long-duration, low-luminosity GRBs such as GRB 060218 have a similar light curve\cite{campana06}. However, the luminosity of \xt\ is 2--3 orders of magnitude lower than that of GRB 060218 and other members of this class. More importantly, like other LGRBs, GRB 060218 resided in a dwarf star-forming galaxy (with $M_*\approx 10^7\ M_\odot$)\cite{wiersema07}, which is very different from the host galaxy of \xt\ that is a main-sequence galaxy (with $M_*\approx 10^9\ M_\odot$) with a relatively low SFR. The large offset of the source from the host galaxy is also at odds with an LGRB origin. 
 \item Shock breakout events such as the X-ray Outburst XRO 080109\cite{soderberg08} have a luminosity 1--2 orders of magnitude lower than that of \xt. The shape of the light curve of XRO 080109 is very different from that of \xt, which shows no evidence of a magnetar-powered plateau. Even though the host galaxy of XRO 080109 (i.e., NGC 2770) is a regular spiral galaxy, the transient occurred in the brightest region of the host galaxy, being consistent with a massive star core collapse origin. Indeed it was associated with a Type Ibc SN 2008D. In contrast, the location of \xt\ is offset from the host galaxy, with little evidence of star formation in the neighborhood. Therefore, a shock breakout origin is disfavored.
\end{itemize}

%% References for Methods
%%
%\vskip 0.5in
%\noindent
%[Here are references in Methods.]

\end{methods}

%% References
%%
%% Put the bibliography here, most people will use BiBTeX in
%% which case the environment below should be replaced with
%% the \bibliography{} command.

%% Use \item's to separate, default label is "Acknowledgements"
%%
\begin{addendum}

 \item We thank A.~Lien, K.~Hurley, and D.~Svinkin for looking into the relevant Swift/BAT, IPN3, and Wind/KNOUS data for us, respectively. 
Y.Q.X., X.C.Z., J.Y.L., and M.Y.S. acknowledge support from the
973 Program (2015CB857004), the National Natural Science
Foundation of China (NSFC-11473026, 11890693, 11421303), the
CAS Frontier Science Key Research Program (QYZDJ-SSW-SLH006),
the CSC (China Scholarship Council)-Leiden University joint scholarship program,
the China Postdoctoral Science Foundation (2016M600485), and the K.C. Wong Education Foundation.
Y.L. acknowledges support by the KIAA-CAS Fellowship, which is jointly supported by Peking University and Chinese Academy of Sciences.
W.N.B. and G.Y. acknowledge support from Chandra X-ray
  Center grant AR8-19016X, NASA grant NNX17AF07G, NASA
  ADP grant 80NSSC18K0878, and the V.M. Willaman Endowment.
B.L. acknowledges support from the National
Key R\&D Program of China (2016YFA0400702) and
NSFC-11673010.
B.-B.Z. acknowledges support from 
National Thousand Young Talents program of China, the National Key Research and Development Program of China (2018YFA0404204), and NSFC-11833003.
F.E.B. acknowledges support from CONICYT-Chile (Basal AFB-170002, FONDECYT Regular 1141218, FONDO ALMA 31160033) and the Ministry of Economy, Development, and Tourism's Millennium Science Initiative through grant IC120009, awarded to The Millennium Institute of Astrophysics, MAS.
F.V. acknowledges financial support from CONICYT and CASSACA through the fourth call for tenders of CAS-CONICYT fund.

 \item[Author Contributions] Y.Q.X. and X.C.Z. discovered \xt, and led the Chandra data reduction and analysis, estimation of rise time and event rate density, and compilation and analysis 
of multiwavelength properties, with inputs from W.N.B., B.L., F.E.B., B.D.L., G.Y., X.K., J.Y.L., M.Y.S., J.-X.W., and F.V.    
B.Z. and X.-F.W. proposed the theoretical model to explain the data. 
Y.Q.X., X.C.Z., and B.Z. wrote the manuscript.
Y.L. analyzed the HST data, measured the offset with inputs from X.C.Z. and Y.Q.X., computed the $F_{\rm light}$ parameter and
the probability O(II:I)$_{\rm host}$, and compared X-ray and gamma-ray properties with that of SGRBs.
B.-B.Z. analyzed the Fermi data and computed relevant flux and luminosity upper limits; B.-B.Z. and X.C.Z. examined the INTEGRAL and Swift data.
H.S. calculated the local event rate density with inputs from X.C.Z. and Y.Q.X.
All authors contributed to the manuscript.

 \item[Author Information] Reprints and permissions information is available 
 at www.nature.com/reprints. The authors declare no competing financial 
 interests. Readers are welcome to comment on the online version of the paper. 
 Correspondence and requests for materials should be addressed to 
 YQX (xuey@ustc.edu.cn), XCZ (zheng@strw.leidenuniv.nl), or BZ (zhang@physics.unlv.edu).

\item[Data Availability] The data that support the plots within this paper 
and other finding of this study are
available from the corresponding authors upon reasonable request.

\item[Competing Interests] The authors declare that they have no competing financial interests.

\end{addendum}

\newcounter{mybibstartvalue}
\setcounter{mybibstartvalue}{30}

\xpatchcmd{\thebibliography}{%
  \usecounter{enumiv}%
}{%
  \usecounter{enumiv}%
  \setcounter{enumiv}{\value{mybibstartvalue}}%
}{}{}

\begin{table}
    \centering
    \caption{Light curve of \xt\ in logarithmic bins}
    \label{tab:loglc}
    \begin{tabular}{rrrrrr}
    \hline
         Time$^\ast$ & Bin width & Counts & Count rate$^\dagger$ & $F_{0.5-7\,\rm keV}$$^\ddagger$ & $L_{\rm 0.3-10\ keV}$$^\ddagger$ \\
         (s) & (s) & & (count s$^{-1}$) & (erg\,s$^{-1}$\,cm$^{-2}$) & (erg\,s$^{-1}$) \\\hline
         4 & 8 & 0 & 0.0 & $0.0^{+2.9}\times 10^{-12}$ & $0.0^{+1.0}\times 10^{46}$ \\
         16 & 16 & 1 & $6.3^{+14.5}_{-5.2}\times 10^{-2}$ & $7.8^{+18.2}_{-6.5}\times 10^{-13}$ & $2.7^{+6.3}_{-2.3}\times 10^{45}$ \\
         40 & 32 & 2 & $6.3^{+8.3}_{-4.0}\times 10^{-2}$ & $7.8^{+10.4}_{-5.0}\times 10^{-13}$ & $2.7^{+3.6}_{-1.8}\times 10^{45}$ \\
         88 & 64 & 3 & $4.7^{+4.6}_{-2.5}\times 10^{-2}$ & $5.9^{+5.7}_{-3.2}\times 10^{-13}$ & $2.0^{+2.0}_{-1.1}\times 10^{45}$\\
         184 & 128 & 3 & $2.3^{+2.3}_{-1.3}\times 10^{-2}$ & $2.9^{+2.9}_{-1.6}\times 10^{-13}$ & $1.0^{+1.0}_{-0.6}\times 10^{45}$\\
         376 & 256 & 10 & $3.9^{+1.7}_{-1.2}\times 10^{-2}$ & $4.9^{+2.1}_{-1.5}\times 10^{-13}$ & $1.7^{+0.7}_{-0.5}\times 10^{45}$\\
         760 & 512 & 15 & $2.9^{+1.0}_{-0.7}\times 10^{-2}$ & $3.7^{+1.2}_{-0.9}\times 10^{-13}$ & $1.3^{+0.4}_{-0.3}\times 10^{45}$\\
         1528 & 1024 & 36 & $3.5^{+0.7}_{-0.6}\times 10^{-2}$ & $4.4^{+0.9}_{-0.7}\times 10^{-13}$ & $1.5^{+0.3}_{-0.3}\times 10^{45}$\\
         3064 & 2048 & 42 & $2.1^{+0.4}_{-0.3}\times 10^{-2}$ & $2.6^{+0.5}_{-0.4}\times 10^{-13}$ & $8.9^{+1.6}_{-1.4}\times 10^{44}$\\
         6136 & 4096 & 13 & $3.2^{+1.2}_{-0.9}\times 10^{-3}$ & $4.0^{+1.4}_{-1.1}\times 10^{-14}$ & $1.4^{+0.5}_{-0.4}\times 10^{44}$\\
         12280 & 8192 & 9 & $1.1^{+0.5}_{-0.4}\times 10^{-3}$ & $1.4^{+0.6}_{-0.4}\times 10^{-14}$ & $4.8^{+2.2}_{-1.6}\times 10^{43}$\\
         24568 & 16384 & 2 & $1.2^{+1.6}_{-0.8}\times 10^{-4}$ & $1.5^{+2.0}_{-1.0}\times 10^{-15}$ & $5.3^{+7.1}_{-3.4}\times 10^{42}$ \\\hline
    \end{tabular}
    \\
    $^\ast$ The time value is the middle time in each bin ($t$=0~s is set to be 10~seconds before the arrival of the first photon).\\
    $^\dagger$ These relatively low count rates essentially eliminate the pile-up issue.\\
    $^\ddagger$ The flux and luminosity values are obtained based on the count rate, exposure time, and overall power-law spectral slope of \xt.
\end{table}

\begin{table*}
    \centering
    \caption{X-ray spectral fitting results for \xt}
    \resizebox{\linewidth}{!}{
    \begin{tabular}{l|l|c|c|c}
    \hline
  Case (Spec\_1 \& Spec\_2) & Spectral fitting results & $C$/d.o.f & goodness & AIC \\\hline
A: free $\Gamma$ \& free $N_{\rm H}$ & $\Gamma_1=1.45^{+0.68}_{-0.50}$, $\Gamma_2=2.67_{-0.77}^{+0.92}$, $N_{\rm H1}=0.45_{-0.45}^{+1.60}$, $N_{\rm H2}=1.03_{-1.03}^{+1.54}$ & 361.61/882 & 53\% & 373.61 \\
B: free $\Gamma$ \& linked $N_{\rm H}$ & $\Gamma_1=1.57_{-0.50}^{+0.55}$, $\Gamma_2=2.53_{-0.64}^{+0.74}$, $N_{\rm H1}=N_{\rm H2}=0.77_{-0.77}^{+1.06}$ & 361.81/883 & 49\% & 371.81\\
C: linked $\Gamma$ \& free $N_{\rm H}$ & $\Gamma_1=\Gamma_2=2.01_{-0.42}^{+0.54}$, $N_{\rm H1}=1.45_{-1.19}^{+1.65}$, $N_{\rm H2}=0.25_{-0.25}^{+0.99}$ & 365.35/883 & 48\% & 375.35 \\
D: linked $\Gamma$ \& linked $N_{\rm H}$ & $\Gamma_1=\Gamma_2=1.93_{-0.46}^{+0.52}$, $N_{\rm H1}=N_{\rm H2}=0.61_{-0.61}^{+1.00}$ & 368.14/884 & 44\% & 376.14 \\\hline
Spec\_0 & $\Gamma=1.93_{-0.47}^{+0.52}$, $N_{\rm H}=0.61_{-0.61}^{+1.00}$ & 261.99/441 & 44\% & --- \\\hline
    \end{tabular}
    }
{\bf Notes.} $\Gamma_1$ and $N_{\rm H1}$ are for Spec\_1; $\Gamma_2$ and $N_{\rm H2}$ are for Spec\_2; $N_{\rm H}$ values are in units of $10^{22}\rm \, cm^{-2}$; d.o.f is the degrees of freedom; all goodness values (obtained using the {\tt goodness} command in {\tt XSPEC}) are around 50\%, which indicates that the fits are good.
\end{table*}

\begin{table*}
    \centering
    \caption{Properties of the 6 closest galaxies to \xt}
    \resizebox{\linewidth}{!}{
    \begin{tabular}{crccccclccrc}
    \hline
    \# &CANDELS & RA & DEC & Offset~($^{\prime\prime}$) & $m_{\rm F606W}$ & $m_{\rm F160W}$ & $z_{\rm best}$ &$M_{\rm F606W}$ & log$(M_{*}/M_{\odot})$ & SFR $(M_{\odot}$ yr$^{-1})$ & Morphology \\
    (1) & (2) &(3)&(4)&(5)&(6)&(7)&(8)&(9)&(10)&(11)&(12)\\
    \hline
    1&4167  & 53.0765804 & $-$27.8733154 & 0.42 & 25.35 & 23.85 & 0.738                  &$-$17.93 &9.07        & 0.81  & Irregular disk  \\
    2&4210  & 53.0760576 & $-$27.8736135 & 1.57 & 23.15 & 21.49 & 0.740              &$-$20.14 &9.77    & 15.49 & Pure bulge  \\
    3&4059  & 53.0765953 & $-$27.8738338 & 1.63 & 25.45 & 24.82 & 3.140 (3.06--3.20) &$-$21.68 &8.83    & 19.50 & N/A  \\
    4&4032  & 53.0766334 & $-$27.8742376 & 3.07 & 25.85 & 24.61 & 1.218 (1.15--1.29) &$-$18.78 &8.75    & 0.52  & N/A \\
    5&28140 & 53.0775843 & $-$27.8727073 & 4.29 & 27.13 & 26.75 & 1.638 (1.49--1.78) &$-$18.28 &8.04    & 0.37  & N/A \\
    6&28134 & 53.0751600 & $-$27.8730197 & 4.42 & 29.00 & 26.88 & 1.688 (0.73--3.15) &$-$16.50 &8.42    & 0.78  & N/A\\
    \hline
    \end{tabular}
    }
{\noindent {\bf Notes.} (1) Object number. (2) CANDELS ID. (3, 4) F160W-band position$^\GuoOneThree$. (5) Nominal offset between the F160W-band position and the X-ray position of XID$_{\rm 7Ms}$~330$^\LuoOneSeven$, without performing a careful astrometric alignment between the two catalogs$^{\GuoOneThree,\LuoOneSeven}$. (6, 7) Apparent AB magnitudes in the F606W and F160W bands. (8) Best redshift estimate$^\SantiniOneFive$. The redshifts of 4167 and 4210 are spectroscopic, while the other four are photometric, with their 1-$\sigma$ confidence ranges shown in parentheses. (9) Absolute F606W-band AB magnitude based on the best redshift. (10, 11) Median stellar mass and SFR of the SED-fitting results from five teams$^\SantiniOneFive$. (12) Morphology measurement$^\HuertasOneFive$.}
    \label{tab:host}
\end{table*}

\begin{figure}[t]
\begin{tabular}{c}
\includegraphics[keepaspectratio, clip, width=1.15\textwidth]{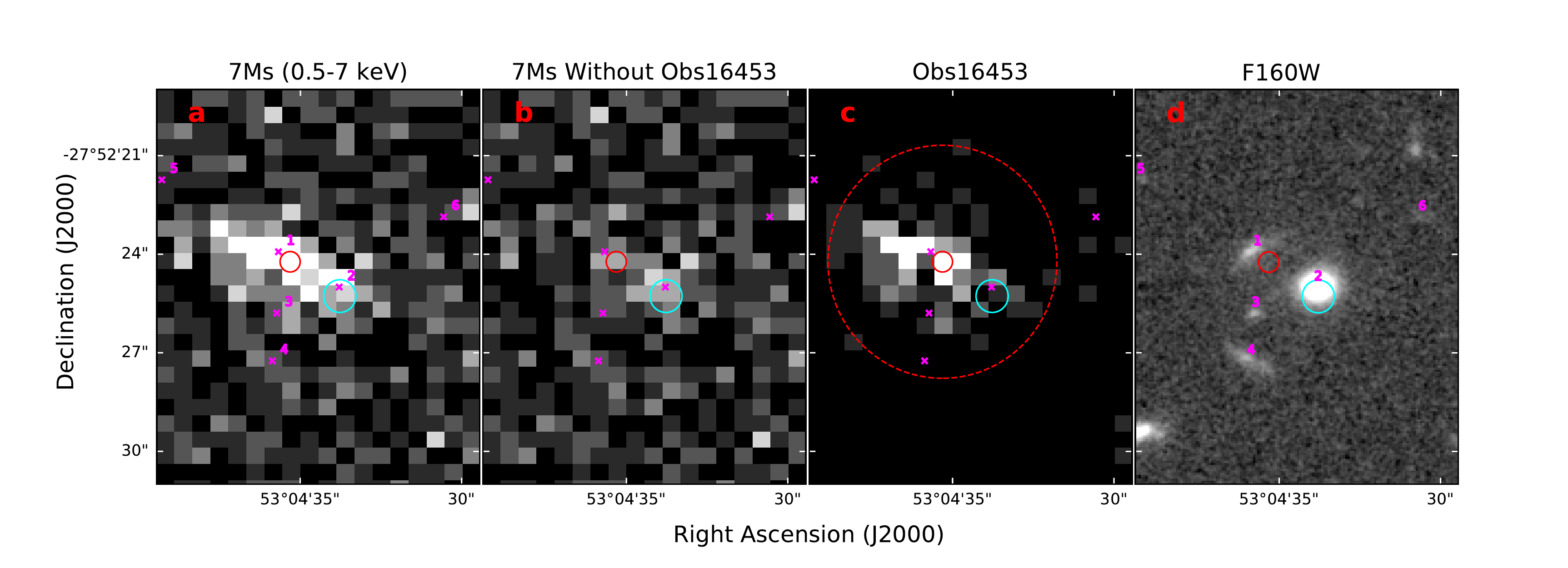} \\
\end{tabular}
\caption{{\bf Multiwavelength images of the \xt\ neighborhood. a,} Merged Chandra \hbox{0.5--7\,keV} image including the entire 7~Ms CDF-S survey. {\bf b,} Same as {\bf a}, but excluding ObsID~16453. {\bf c,} Chandra \hbox{0.5--7\,keV} image from ObsID~16453 alone. The large dashed circle marks the source region ($r=3.5^{\prime\prime}$) for the extraction of light curves and spectra. The X-ray images in {\bf a}--{\bf c} are rendered in counts using the same linear scale, with pixel values ranging from 0 to 8. {\bf d,} HST/CANDELS F160W image.
In each panel, the small red circle marks the position and 1~$\sigma$ positional uncertainty of the source with \src reported in the 7~Ms CDF-S main catalog$^\LuoOneSeven$, while the cyan circle denotes that of the source with XID$_{\rm 4Ms}$~256 reported in the 4~Ms CDF-S main catalog$^\XueOneOne$.
The magenta numbers and crosses mark the object numbers of the 6 closest galaxies to \xt\ (see Table~2) and their positions in the CANDELS F160W DR1 catalog$^\GuoOneThree$, respectively.
For clarity, the position of \xt\ is not annotated, which is slightly ($\approx0.3^{\prime\prime}$) northeastern to that of XID$_{\rm 7Ms}$~330.}
\label{fig:host}
\end{figure}

\end{document}